\documentclass[prc,pdftex]{revtex4}

\usepackage{graphicx,amsmath}
\begin{document}
\title{Alternative Contributions to the Angular Correlations Observed at RHIC Associated with Parity Fluctuations}
\author{Scott Pratt}
\affiliation{Department of Physics and Astronomy and National Superconducting Cyclotron Laboratory,
Michigan State University\\
East Lansing, Michigan 48824}
\date{\today}

\begin{abstract}
Recent measurements at RHIC of angular correlations of same-sign vs. opposite sign pairs have been interpreted as evidence for large-scale fluctuations of parity-odd fields. In this paper, we provide alternative explanations of the same phenomena based on correlations from charge and momentum conservation overlaid with elliptic flow. These effects are shown to produce correlations with similar magnitudes as those measured. Other correlations are also considered, but estimates of their size suggest they are inconsequential. 
\end{abstract}

\maketitle

\section{Introduction}

The prospect of observing large fluctuations of parity-odd observables in heavy ion collisions has recently gained great attention due to measurements by the STAR collaboration at the RHIC (The Relativistic Heavy Ion Collider) \cite{STARparity}. It has been proposed that the fluctuations derive from the color flux tubes, which carry both color-electric and color-magnetic flux parallel to the beam direction for the first $\sim 1/2$ fm/$c$ of the collision \cite{Kharzeev:2004ey,Kharzeev:2009mf}. The magnetic field lines, which have $\nabla\cdot{\bf B}^a\ne 0$ in QCD, can be randomly parallel or anti-parallel to the electric field lines. However, since each flux tube might be responsible for dozens of particles, there might exist the opportunity to view these fluctuations. Observation of such fluctuations would validate the existence of coherent color magnetic flux, which is a basic feature of a non-Abelian gauge theory.

The experimental manifestation of fluctuations comes from the parity-odd value of ${\bf E}^a\cdot{\bf B}^a$, which is randomly positive or negative within a given tube. Via anomalous chiral couplings, these fields couple to the electromagnetic field ${\bf E}\cdot{\bf B}$. In the participant region of a heavy ion collision, where there exists a strong coherent transverse magnetic field ${\bf B}$ due to the passing ions, one can then generate a non-zero electric field due to the coupling with the color fields. This electric field is then randomly parallel or anti-parallel to the electric field, and should last a few tenths of a fm/$c$. Since each flux tube generates an electric field with random signs, the averaged electric field fluctuates as $1/\sqrt{N_{\rm tubes}}$, where $N_{\rm tubes}$ is the number of such tubes. Angular correlations might then scale as $1/N_{\rm tubes}$. Since there are fewer tubes than particles, one might hope that this fluctuation stands out compared to random fluctuations. 

The size of the correlations observed by STAR is of the order of $10^{-3}$ to $10^{-4}$, and given that these typically have multiplicities, $M$, of a few hundred particles at mid-rapidity, the correlations are smaller than 
$1/M$ and much smaller than $1/N_{\rm tubes}$, so one cannot neglect correlations that involve a handful of particles, such as charge or momentum conservation. The manifestations of charge and momentum conservation are discussed here, and are shown to provide effects which should explain the bulk, though perhaps not all, of the observed correlation. Other contributions are also discussed, but are found to be significantly smaller than the observed signal. 

\begin{figure}
\centerline{\includegraphics[width=0.5\textwidth]{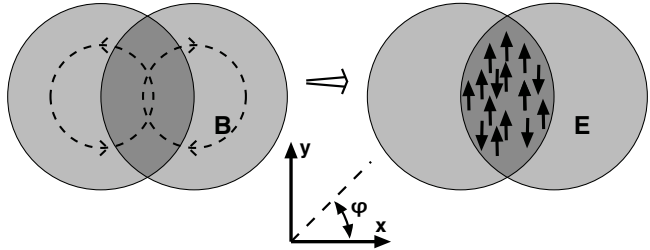}}
\caption{\label{fig:paritycartoon}
The overlap of two heavy ions from the perspective of looking down the beam line. In the left panel, one sees the magnetic field lines from the two ions, which add constructively in the participant (dark grey) region. These field lines couple to the ${\bf E}^a\cdot{\bf B}^a$ values which fluctuate randomly within the individual color flux tubes, to make electric fields illustrated in the right-hand panel. The signs of the ${\bf E}$ fields are of the same sign within a given flux tube, which extends over a large range of rapidities, but when averaged over the entire participant region, largely cancel and leave an average field strength lessened by a factor of $1/\sqrt{N_{\rm tubes}}$.}
\end{figure}
The parity-odd observable measured by STAR is the quantity,
\begin{eqnarray}
\gamma_{\alpha\beta}&\equiv&\langle \cos(\phi_\alpha+\phi_\beta)\rangle\\
\nonumber
&=&\frac{\sum_{i\in\alpha,j\in\beta}\cos(\phi_i+\phi_j)}{M_\alpha M_\beta},
\end{eqnarray}
where $\alpha$ and $\beta$ represent either the positive or negative charges, and $M_\alpha$ and $M_\beta$ are the multiplicities of each type per event. If the two types are identical, the $i=j$ term is omitted from the sum. The sum is calculated in each event, then averaged over many events. The angles $\phi_i$ are measured relative to the reaction plane of the event, as pictured in Fig. \ref{fig:paritycartoon}.
The observable $\gamma_{\alpha\beta}$ might have been defined as simply $\langle\sin\phi_\alpha\sin\phi_\beta\rangle$, rather than with $\langle\cos(\phi_\alpha+\phi_\beta)\rangle=\langle\cos\phi_\alpha\cos\phi_\beta-\sin\phi_\alpha\sin\phi_\beta\rangle$. By taking the difference with the $\cos\phi_\alpha\cos\phi_\beta$ terms, it would be expected that certain contributions, such as those from resonances, might be subtracted away.

\begin{figure}
\centerline{\includegraphics[width=0.5\textwidth]{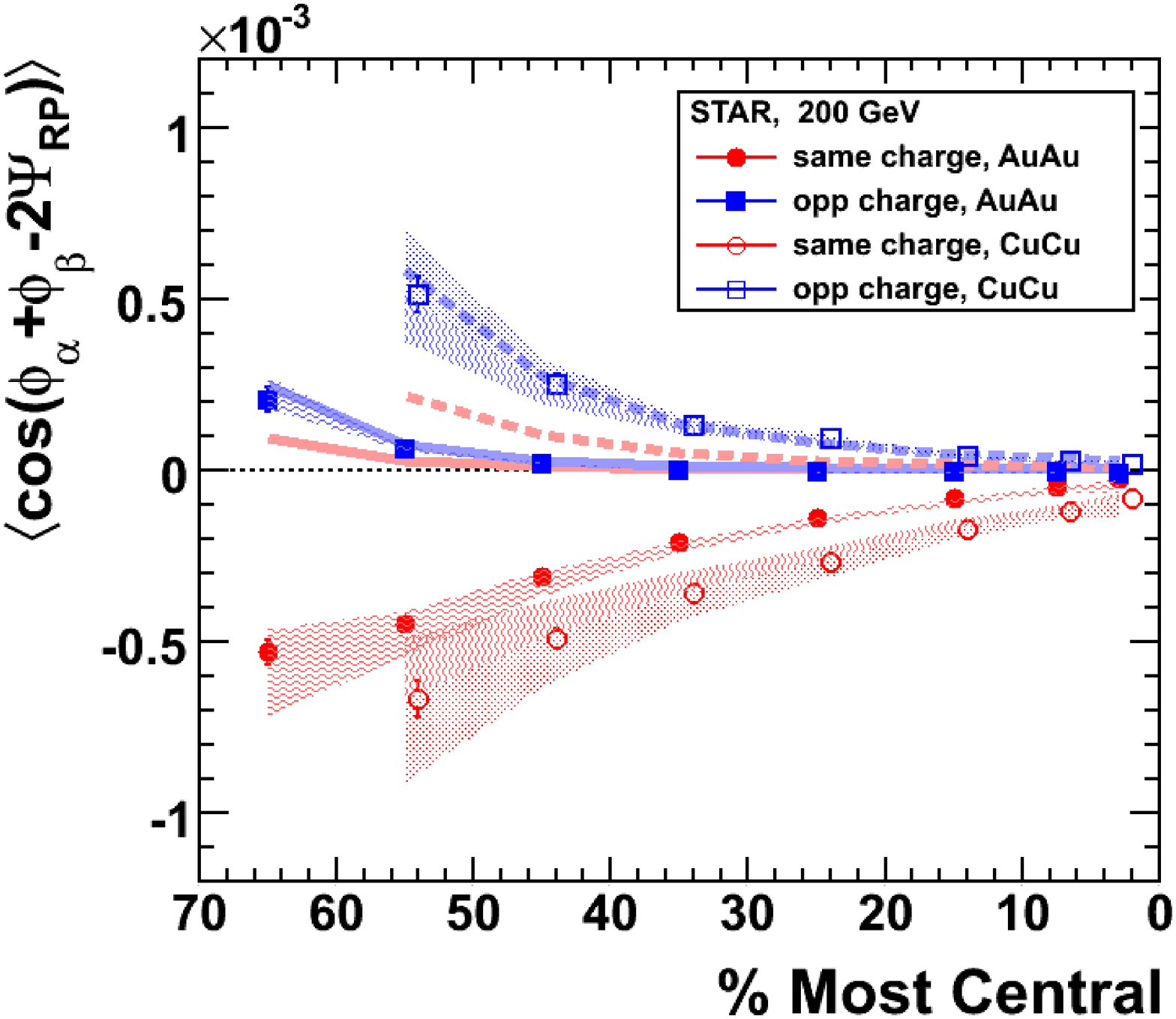}}
\caption{\label{fig:starprl}
Angular correlations as measured by STAR for both Au+Au and Cu+Cu collisions. The shaded areas reflect systematic uncertainties in the analysis related to the elliptic flow.
}
\end{figure}

STAR's measurements for both same-sign and opposite-sign correlations are shown in Fig. \ref{fig:starprl} for Au+Au and Cu+Cu collisions with beams of $100A$ GeV. If the source of the correlation was solely the induced electric field, the opposite-sign correlation, $\gamma_{+-}$ would have been positive and the same-sign correlation, $\gamma_{ss}$ would have been equal and opposite. Indeed, $\gamma_{ss}$ is lower than $\gamma_{+-}$, though they are not equal and opposite. The correlations fall with multiplicity, as one would expect for all such correlations. Since phenomena such as jets or momentum conservation might affect both same-sign and opposite-sign pairs, it seems reasonable to focus on the difference, $\gamma_{+-}-\gamma_{ss}$, as well as individual quantities.

The next two sections focus on contributions from charge conservation and momentum conservation respectively, with ther possible contributions being discussed in the subsequent section. Quantitative comparisons are presented in the final section along with conclusions.

\section{Charge conservation}\label{sec:chargeconservation}

Whenever a positive charge is created, a negative charge arises from the same point in space time. Given the short mean free paths of particles in heavy ion collisions, both particles then tend to be focused in the same rapidity and azimuthal angle by collective flow. This results in a correlation between positive and negative particles, i.e., for every positive particle emitted at an angle $\phi_+$, there tends to be a negative particle emitted with $\phi_-\approx \phi_+$ and with similar rapidity. Charge balance functions \cite{Adams:2003kg,Bass:2000az,Cheng:2004zy} represent a measure of such correlations, and have already been measured as a function of relative rapidity for identified particles and for relative pseudo-rapidity $\eta$ for non-identified particles. Charge balance functions are calculated by taking like-sign subtractions of positive and negative particles,
\begin{equation}
B(\Delta\eta)\equiv\frac{N_{+-}(\Delta\eta)+N_{-+}(\Delta\eta)-N_{++}(\Delta\eta)-N_{--}(\Delta\eta)}{(N_++N_-)},
\end{equation}
where $N_{ab}$ is the number of pairs per event separated by relative rapidity $\Delta\eta$. The separation of the balancing charges is consistent with the charges being emitted from regions with the same collective rapidity, with the separation being solely driven by the final thermal motion of the decoupling source. This spread tends to be of the order of a half unit of rapidity. Similarly, one would expect that any charged particle would be accompanied by a balancing charge with a similar relative angle. The degree of correlation can be quantified by $\langle\cos(\Delta\Phi_{\rm balance})\rangle$, which would be unity if the balancing charge was always emitted exactly along the same direction, and would be zero if the balancing charge were to be emitted randomly. Such preliminary correlations have been presented by STAR \cite{Westfall:PC}, and show that the balancing charge tends to be emitted in within a few tens of degrees of its companion in central collisions. As expected, the correlations are significantly broader in relative angle for peripheral collisions.

For the case of correlations to be considered here,
\begin{eqnarray}
\label{eq:gammapm}
\gamma_{+-}&=&\frac{\sum_{i\in +,j\in -}\cos(2\phi_i+(\phi_j-\phi_i))}{M_+M_-}\\
\nonumber
&=&f_Q\frac{\sum_i\left(\cos 2\phi_i\langle\cos\Delta\phi_{\rm balance}\rangle(\phi_i)-\sin 2\phi_i\langle\sin\Delta\phi_{\rm balance}\rangle(\phi_i)\right)}{M_+},
\end{eqnarray}
where it has been assumed that only one charge $j$ balances the charge $i$ and that its relative angle $\Delta\phi_{\rm balance}=\phi_j-\phi_i$. Here, $f_Q$ is the fraction of the charge that will be found in the sum $\sum_j$, and can be less than unity due to the finite acceptance and efficiency of the detector.

If the balancing charge were always emitted in the same direction as $\phi_i$, one could replace $\langle\cos\Delta\phi_{\rm balance}\rangle=1$ and ignore the second term in Eq. (\ref{eq:gammapm}). If $\langle\cos\Delta\phi_{\rm balance}\rangle$ were assumed to be independent of the angle $\phi_i$, and if the balancing charge were equally like to be found at positive and negative $\Delta\phi_{\rm balance}$, Eq. (\ref{eq:gammapm}) simplifies to
\begin{equation}
\label{eq:gamma_chargeconservation}
\gamma_{+-}=2f_Q\frac{v_2}{M}\langle\langle\cos\Delta\phi_{\rm balance}\rangle\rangle,
\end{equation}
where $M=M_++M_-$, and assuming $M_+\approx M_-$. The double brackets emphasize that the average is also over the direction of $\phi$. More generally, the distributions of $\Delta\phi_{\rm balance}$ might depend on the direction of $\phi_i$. Keeping up to second order harmonics,
\begin{eqnarray}
\langle\cos\Delta\phi_{\rm balance}\rangle(\phi_i)&=&\langle\langle\cos\Delta\phi_{\rm balance}\rangle\rangle
+2v_{2c}^{(B)}\cos 2\phi_i,\\
\nonumber
\langle\sin\Delta\phi_{\rm balance}\rangle(\phi_i)&=&2v_{2s}^{(B)}\sin 2\phi_i,\\
\nonumber
v_{2c}^{(B)}&\equiv& \frac{1}{2\pi}\int d\phi_i \langle\cos\Delta\phi_{\rm balance}\rangle(\phi_i)\cos 2\phi_i,\\
\nonumber
v_{2s}^{(B)}&\equiv& \frac{1}{2\pi}\int d\phi_i \langle\sin\Delta\phi_{\rm balance}\rangle(\phi_i)\sin 2\phi_i.
\end{eqnarray}
With these definitions, the general form of Eq. (\ref{eq:gammapm}) becomes
\begin{equation}
\label{eq:gamma_chargeconservation_general}
\gamma_{+-}=\frac{2f_Q}{M}\left( v_2\langle\langle\cos\Delta\phi_{\rm balance}\rangle\rangle
+v_{2c}^{(B)}-v_{2s}^{(B)}\right).
\end{equation}
Since the in-plane charge-balance pairs are expected to be more tightly correlated in angle, due to more collective flow and less surface curvature for in-plane emission, $v_{2c}^{(B)}$ should be positive. For $\phi_i$ at 45 degrees, one expects the balancing charge to have a slight preference to be emitted at a smaller angle due to elliptic flow which gives more particles for $\phi=0$ than for $\phi=90$ degrees. This suggests that $v_{2s}^{(B)}$ would be negative, which would also give a positive contribution to $\gamma_{+-}$ in Eq. (\ref{eq:gamma_chargeconservation_general}). Assuming that the balancing charges are highly correlated, i.e. $\langle\langle\cos\Delta\phi_{\rm balance}\rangle\rangle\lesssim 1$, there is little opportunity for the distribution of balancing angles to vary with $\phi_i$ sufficiently to make $v_{2c}^{(B)}$ or $v_{2s}^{(B)}$ very large. Assuming $\langle\langle\cos\Delta\phi_{\rm balance}\rangle\rangle$ is fairly large, the first term in Eq. (\ref{eq:gamma_chargeconservation_general}) should dominate the other two terms.

The fraction $f_Q$ can be related to the normalization of the charge balance function, which is near 0.5 in STAR measurements. About half of the reduction comes from the fact that some particles within the $\pm 1$ unit of rapidities covered by STAR are not recorded due to finite efficiency or for having too low of a transverse momentum $p_t$. The other half of the reduction comes from the balancing charge lying outside the rapidity range of the detector. For a particle with rapidity of zero, the balancing charge tends to be largely inside the acceptance, while for a particle emitted barely within the rapidity acceptance, the balancing charge has a 50\% chance of being outside. Thus, overall the contribution from $f_Q$ due to the balancing charge being outside the acceptance is about 0.75. If one were to use the efficiency and acceptance corrected values of the multiplicity in Eq. (\ref{eq:gamma_chargeconservation}), rather than the number $M$ of charges actually recorded, the fraction $f_Q$ one would use should be on the order of 0.75. Given that collective flow accounts for about half of the final motion of the particles, one might expect $\cos\Delta\Phi_{\rm balance}\sim 0.5$. For very peripheral collisions, one would expect a smaller value. If the charge balance function were measured fully as a function of $\phi_\alpha$ and $\phi_\beta$, all of the quantities on the right-hand side of Eq. (\ref{eq:gamma_chargeconservation}) can be extracted from experimental analyses.

\section{Momentum Conservation}\label{sec:momentumconservation}

Ignoring charge conservation for the moment, the implications of momentum conservation are best understood by considering the correlation calculated independently of the charge of the particle. In that case,
\begin{equation}
\label{eq:gammadef}
\gamma\equiv \frac{\sum_{i\ne j}\cos\left(\phi_i+\phi_j\right)}{M(M-1)}=\frac{\sum_{i\ne j} \left(\cos\phi_i\cos\phi_j-\sin\phi_i\sin\phi_j\right)}{(M_{\rm tot})(M_{\rm tot}-1)},
\end{equation}
where $i$ and $j$ are summed over both charged and neutral particles, i.e, from 1 to $M_{\rm tot}=M_++M_0+M_-$. Momentum conservation implies,
\begin{equation}
\sum_i p^{(i)}_x=\sum_ip^{(i)}_y=0,
\end{equation}
The effects of momentum conservation would be mostly model-independent if the definition of $\gamma$ was defined in terms of $p_x^{(i)}=p_t^{(i)}\cos\phi_i$ instead of the angles alone. Instead, we will make a simplifying assumption that all particles have the same $p_t$. This approximation should be good enough to estimate the effects of momentum conservation to within a few tens of percent. With this assumption,
\begin{equation}
\sum_i\cos\phi_i=\sum_i\sin\phi_i=0.
\end{equation}
Using this constraint in Eq. (\ref{eq:gammadef}),
\begin{equation}
\gamma=-f_P\frac{\sum_i \left(\cos^2\phi_i-\sin^2\phi_i\right)}{M_{\rm tot}^2},
\end{equation}
where $f_P$ is the fraction of the momentum balance found within the acceptance, i.e.,
\begin{equation}
\sum_{j\ne i} \cos\phi_i=-f_P\cos\phi_i.
\end{equation}
Again, using the definition of $v_2$,
\begin{equation}
\gamma=-f_P\frac{v_2}{M_{\rm tot}},
\end{equation}
Thus, the correlation $\gamma$ is generated by a combination of momentum conservation, which causes particles to be preferably generated in the opposite direction and elliptic flow, which gives more particles in the $\pm x$ direction than in the $\pm y$ direction.

Experimentally, the same-sign is defined as the average of $\gamma_{++}$ and $\gamma_{--}$, and assuming that the momentum balance is shared equally amongst the charges
\begin{equation}
\label{eq:gamma_momentumconservation}
\gamma_{ss}=\frac{1}{2}(\gamma_{++}+\gamma_{--})=-f_P\frac{v_2}{M_{\rm tot}}.
\end{equation}

As stated above, this result was predicated on the assumption that all particles had the same $p_t$. To remove this model dependence, one could consider the same quantity weighted with the transverse momenta of the particles,
\begin{equation}
\label{eq:gammaprimedef}
\gamma^\prime\equiv \frac{\sum_{i\ne j} p_t^{(i)}p_t^{(j)}\cos(\phi_i+\phi_j)}{M_{\rm tot}\langle p_t^2\rangle}
=\frac{\sum_{i\ne j}(p_x^{(i)}p_x^{(j)}-p_y^{(i)}p_y^{(j)})}{M_{\rm tot}\langle p_t^2\rangle},
\end{equation}
with the sums covering all charged particles. In that case, one comes up with the same answer as in Eq. (\ref{eq:gamma_momentumconservation}), but without the model dependence, and with $v^\prime_2$ being weighted with $p_t^2$,
\begin{equation}
\gamma^\prime=-f_P\frac{v_2^\prime}{M_{\rm tot}}\left(1+\langle\cos\Phi_{\rm balance}\rangle\right), ~~v'_2\equiv \frac{\sum_i p_{i,t}^2\cos 2\phi_i}{M_{\rm tot}\langle p_t^2\rangle}.
\end{equation}

In practice, only a subset of particles are measured. In that case some of the momentum balance comes from unmeasured particles and one might expect $-\gamma_{ss}<<v_2/M$. In the STAR experiment at RHIC for which these correlations were measured, tracks were measured for the central two units of rapidity. However, the initial colliding beams approached with $\pm 5.5$ units of rapidity with over half of the tracks emerging with rapidities outside the STAR acceptance. These particles can serve as a source of momentum, which can quench the momentum conservation condition, thus reduce the magnitude of $\gamma_{ss}$. However, the transverse momentum of a given track is more likely to be balanced by neighboring particles, which have similar rapidities. This is particularly true when considering the components of the momenta responsible for elliptic flow. For instance, consider a particle with final momentum transverse momentum $\vec{p}$. Some of that momentum is generated by the passing nuclei, which impart a momentum $\vec{k}$ which is balanced by particles well outside the acceptance. A second component of the momentum $\vec{q}$ comes from interactions with neighbors with similar rapidity. For the track $i$,
\begin{equation}
\vec{p}_i=\vec{k}_i+\vec{q}_i, 
\end{equation}
with $\sum_i\vec{q}_i=0$. Since the momenta $\vec{k}$ are generated at very early times, before particles can gain awareness of the elliptic anisotropy of the collision region, the vectors $\vec{k}_i$ are isotropically distributed in the $xy$ plane. When one calculates the correlation,
\begin{equation}
\gamma^\prime=\frac{\sum_{i\ne j}(k_x^{(i)}k_x^{(j)}-k_y^{(i)}k_y^{(j)})}{M_{\rm tot}\langle p_t^2\rangle}
+2 \frac{\sum_{i\ne j}(k_x^{(i)}q_x^{(j)}-k_y^{(i)}q_y^{(j)})}{M_{\rm tot}\langle p_t^2\rangle}
+ \frac{\sum_{i\ne j}(q_x^{(i)}q_x^{(j)}-q_y^{(i)}q_y^{(j)})}{M_{\rm tot}\langle p_t^2\rangle},
\end{equation}
and given that the vectors $\vec{k}$ are isotropic and uncorrelated,
\begin{equation}
\gamma^\prime=2 \frac{\sum_{i\ne j}(k_x^{(i)}q_x^{(j)}-k_y^{(i)}q_y^{(j)})}{M_{\rm tot}\langle p_t^2\rangle}
+ \frac{\sum_{i\ne j}(q_x^{(i)}q_x^{(j)}-q_y^{(i)}q_y^{(j)})}{M_{\rm tot}\langle p_t^2\rangle}.
\end{equation}
Applying momentum conservation, $\sum_i\vec{q}_i=0$,
\begin{equation}
\gamma^\prime=-2\frac{\sum_{i}(k_x^{(i)}q_x^{(i)}-k_y^{(i)}q_y^{(i)})}{M_{\rm tot}\langle p_t^2\rangle}
- \frac{\sum_{i}(q_x^{(i)2}-q_y^{(i)2})}{M_{\rm tot}\langle p_t^2\rangle}.
\end{equation}
Similarly, one can calculate $v_2$ with the same decomposition and find,
\begin{eqnarray}
\label{eq:v2deriv}
v'_2&=&\frac{\sum_{i}(k_x^{(i)2}-k_y^{(i)2})}{M_{\rm tot}\langle p_t^2\rangle}
+2\frac{\sum_{i}(k_x^{(i)}q_x^{(i)}-k_y^{(i)}q_y^{(i)})}{M_{\rm tot}\langle p_t^2\rangle}
+\frac{\sum_{i}(q_x^{(i)2}-q_y^{(i)2})}{M_{\rm tot}\langle p_t^2\rangle}\\
\nonumber
&=&2\frac{\sum_{i}(k_x^{(i)}q_x^{(i)}-k_y^{(i)}q_y^{(i)})}{M_{\rm tot}\langle p_t^2\rangle}
+\frac{\sum_{i}(q_x^{(i)2}-q_y^{(i)2})}{M_{\rm tot}\langle p_t^2\rangle},
\end{eqnarray}
using the isotropy of $\vec{k}$ to eliminate the first term in the first line of Eq. (\ref{eq:v2deriv}). By inspection of the last two expressions, one again finds
\begin{equation}
\gamma^\prime=\frac{v^\prime_2}{M_{\rm tot}}.
\end{equation}
Thus, since elliptic flow is expected to be generated from local interactions, one expects the damping of elliptic flow from finite acceptance to be due to edge effects. If momentum conservation is confined to within a half unit of rapidity, the damping might be a few tens of percent at most for a large acceptance experiment like STAR.

If one were to explore an observable without the $\cos(\phi_i+\phi_j)$ weight, one would again have the negative contribution from momentum conservation, but it would be more strongly damped by the conservation being spread out over a much greater rapidity range. Furthermore, such an observable is significantly affected by jets or hot spots which make positive contributions to the correlations. Such observables are studied under the moniker of $p_t$ fluctuations.

The contribution from charge correlation was ignored in the expression above for $\gamma_{\rm ss}$, as if the two sources of correlation could be treated independently. However, this is not the case. For a charged particle observed with momentum ${\bf p}$, one expects a particle of similar momentum but opposite charge. The same-sign correlation explicitly ignores the contribution of this balancing particle, but its momentum nonetheless needs to be balanced, and since the momentum balance was assumed to be spread equally amongst all charges, it magnifies $\gamma_{\rm ss}$. The momentum that must be balanced is increased by the factor $(1+\langle\langle\cos\Delta\phi_{\rm balance}\rangle\rangle)$. This gives the final estimate for the same-sign angular correlation,
\begin{equation}
\gamma_{\rm ss}\approx -\frac{2f_P}{3}\frac{v_2}{M}\left(1+\langle\cos\Delta\phi_{\rm balance}\rangle\right),
\end{equation}
where the factor of $2/3$ came from assuming the charged multiplicity is two thirds the total multiplicity, $M=2M_{\rm tot}/3$. The same enhancement factor should also be applied to the expression for the $p_t$ weighted correlation,
\begin{eqnarray}
\label{eq:gammaprime}
\gamma'_{\rm ss}&=&-\frac{2f_P}{3}\frac{v'_2}{M}\left(1+\langle\langle\cos\Delta\phi_{\rm balance}\rangle\rangle\right),\\
\nonumber
v'_2&\equiv& \frac{\sum_{i=1,M}p_{t,i}^2\cos 2\phi_i}{M_{\rm ch}\langle p_t^2\rangle}.
\end{eqnarray}

\section{Other Possible Sources of Angular Correlation}\label{sec:other}

The first source of correlation we consider derives from the electric field of the incoming nuclei, which can contribute to charge separation, i.e., $\gamma_{+-}-\gamma_{ss}$. As illustrated in Fig. \ref{fig:paritycartoon}, the magnetic field lines in the participant region combine constructively, whereas the electric field lines from the two nuclei would average to zero in the participant region. However, the location of the participants within that volume fluctuates given the random nature of the nucleon-nucleon collisions. Once the center-of-mass of the initial collisions strays from the center ($x=y=0$) the averaged electric field is no longer zero. The highly Lorentz-contracted fields from the spectators impart an impulse $\Delta p_y$ to charged particles in the participant region. Due to the contraction the field decays within a few tenths of a fm/$c$, so only the charge from the initial state is affected. For the purposes of making a crude estimate here, we label that fraction $F_0$, and note that it might be at most a few tens of percent. The impulse such a charge $i$ feels from a highly relativistic particle $j$ is
\begin{equation}
\Delta p_y^{(ij)}=2\alpha\frac{(y_i-y_j)}{r^2}.
\end{equation}
To estimate the effect, we considered spectators and participants to be randomly positioned in spheres of 7 fm radius, then using a 35 mb cross section considered the net impulse imparted onto positive charges located according to the collision points from all the protons in the initial beams, which were assumed to move along straight-line trajectories. The rms $\Delta p_y$ for a charge in the participant region was $\sim$ 2.5 MeV/$c$ for central collisions and rose to $\sim$ 5 MeV/$c$ for very peripheral collisions. The induced $\langle\sin\phi\rangle$ for particles of a given charge, given the original distribution of $\phi$ was random, is 
\begin{equation}
\langle\sin\phi\rangle\approx\frac{F_0}{2}\frac{\Delta p_y}{\langle p_t\rangle}.
\end{equation}
The contribution to the angular correlations is then,
\begin{equation}
\gamma_{\rm ss}-\gamma_{+-}=\langle\sin\phi\rangle^2\approx\frac{F_0^2}{2}\frac{\Delta p_y^2}{\langle p_t\rangle^2}~.
\end{equation}
Since $\Delta p_y$ is less than one percent of the mean $p_t$, this effect comes out on the order of $10^{-5}$, which is significantly smaller than the charge- and momentum-conservation effects described in the previous sections. However, for the most central collisions, where the conservation induced effects become very small, there would appear to be a chance that these effects could contribute.

Although rather small, the expression above still significantly over-estimates the effect. This is because the initial impulse will be reduced when the initial momenta are absorbed and thermalized by the medium. Assuming the particles move for a time $\tau_0$ before being absorbed, the charge separation in momentum space translates into a small dipole moment in the thermalized fireball. One can estimate the dipole moment by calculating how far the particles move before $\tau_0$, which might be a few tenths of a fm/$c$. Once that dipole moment is super-imposed onto a simple blast-wave picture of the final state, it again translates into a momentum anisotropy, but tends the value for $\langle\sin\phi\rangle$ tends to be lower by a factor $\tau_0 p_t u_{\perp,{\rm max}}/(TR)$ than the expression above, where $u_{\perp,{\rm max}}$ is the collective velocity at the edge of the fireball ($\sim 1$), $T$ is the decoupling temperature, and $R$ is the initial size of the overlap region. This effectively reduces the effect by an order of magnitude. However, for higher $p_t$ particles, one might expect the particles to avoid thermalization. Furthermore, for higher $p_t$ the fraction $F_0$ might be higher. Thus, it might be expected that a modest, at best, contribution from this effect might survive for higher $p_t$.

The contribution from fluctuating initial conditions should behave very similarly to the effect from parity fluctuations. The fluctuating initial conditions lead to electric fields which are on the order of 10\% of the magnetic fields. It is more difficult to estimate the degree to which electric fields are generated by the anomalous coupling to the QCD fields and the magnetic field. Given that the coupling involves extra powers of both the QCD and electromagnetic coupling constants, the generated electric fields might well be smaller than those considered here. If such is the case, the effect of parity fluctuations would be far too small for observation.

The most direct way to distinguish between the contributions from charge- and momentum- conservation from contributions from fluctuating initial conditions, or equivalently from parity fluctuations, would be to consider the rapidity range over which the correlations extend. For the conservation effects, the correlation should be confined to approximately on unit of relative rapidity, whereas the effect of fluctuation initial conditions should extend several units of rapidity.

In addition to the charge separation due to the Coulomb force, one might also consider forces from the nuclear mean field, which due to the difference in the proton and neutron number would also be non-zero. Although the coupling constant for nuclear forces is larger by two orders of magnitude, the force is short range and only involves a few neighbors. The net result is difficult to predict, but might have a stronger impact than the electric force considered here. Again, the correlation might extend several units of rapidity, but the magnitude of the effect would scale differently with the centrality.

A final source of correlation has to do with the two-particle correlation induced by final-state interactions. These correlations are driven by identical-particle interference, mutual Coulomb attraction or repulsion, or strong interactions between the particles. Strong interactions mainly contribute through resonant interactions, and thus mainly affect opposite sign correlations. However, these contributions are already included as part of the charge balance functions. A decaying neutral particle contributes to the charge balance function by having the balancing charge located in a narrow range of invariant relative momentum, which, depending on the invariant mass, might either broaden or narrow the charge balance function, but not change the overall normalization. Similarly, Coulomb interactions pull positives and negatives toward one another in momentum space, while pushing the positives and negatives further apart. Again, these forces do not change the normalization of the balance function, but can affect the width, leading to a larger value of $\langle\langle\cos\Delta\phi_{\rm balance}\rangle\rangle$. Identical particle statistics are mainly important for pions due to their larger phase space density. This draws same-sign pairs together in momentum space. Again, such correlations cannot change the fact that electric charge is locally conserved, so the normalization of the balance function is unaffected. Since the integrated strength of identical particle correlations is the average phase space-density, which averages to less than 10\% over the acceptance, this implies that less than 10\% of the strength of the balance function is pushed away from very small relative angle to larger relative angles, which might reduce $\langle\langle\cos\Delta\phi_{\rm balance}\rangle\rangle$ by a few percent.

It appears there are few effects that could modify the difference, $\gamma_{+-}-\gamma_{ss}$. From the considerations in this section, it would seem that the combination $\gamma_{+-}-\gamma_{ss}$ should be dominated by charge conservation effects as described by Eq. (\ref{eq:gamma_chargeconservation}), since competing effects would appear to be much smaller. Thus, this expression should be trustworthy to within the 10-20\% level. It is easier to imagine effects that would contribute to $\gamma_{ss}$ besides momentum conservation. The effects of jets being quenched differently between in-plane and out-of-plane is difficult to assess.

\section{Comparisons with Data and Conclusions}
\begin{figure}
\centerline{\includegraphics[width=0.5\textwidth]{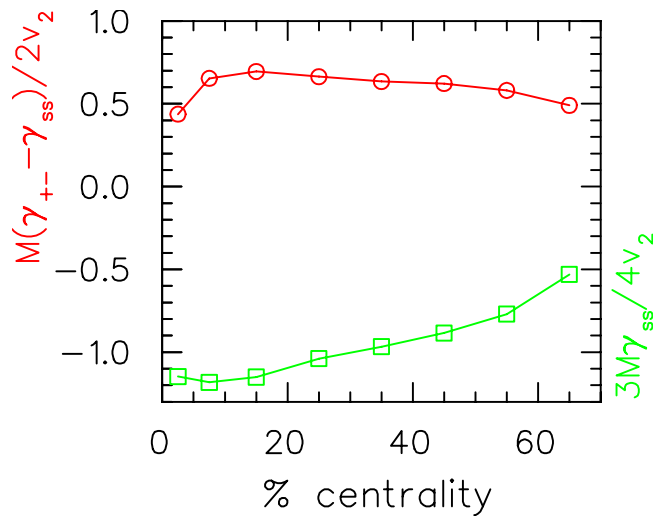}}
\caption{\label{fig:results}
Angular correlations from STAR scaled by the multiplicity and $v_2$. The correlations sensitive to charge correlation, $(\gamma_{+-}-\gamma_{ss})M/2v_2$, are shown by red circles. If the angular width of the balance function is narrowly peaked, and if an observed particle has a high probability of having its charge-balancing partner fall within the rapidity range of the detector, the scaled correlation should be $\lesssim$ unity as shown in Eq. (\ref{eq:scaled}). Thus, it appears there is a possibility that $\gamma_{+-}-\gamma_{ss}$ is largely caused by elliptic flow combining with local charge conservation as described in Sec. \ref{sec:chargeconservation}. The scaled same-sign correlations, $3M\gamma_{ss}/4v_2$, are represented by green squares. Based on the simplifying assumption that all particles have the same $p_t$, Eq. (\ref{eq:scaled}) suggests that these correlations should be $\gtrsim -1$ if solely driven by momentum conservation. The impact of the assumption of equal $p_t$ can be tested by evaluating $\gamma'_{ss}$ described in Eq. (\ref{eq:gammaprime}).
}
\end{figure}

Here, we replot the correlations measured by STAR in a way that better emphasizes how they might compare to correlations induced by momentum and charge conservation. The correlations $\gamma_{+-}-\gamma_{ss}$ and $\gamma_{ss}$ are each presented in Fig. \ref{fig:results} after being multiplied by the multiplicity within the detector and divided by $v_2$. The expressions for the correlation then become,
\begin{eqnarray}
\label{eq:scaled}
\frac{M}{2v_2}\left(\gamma_{+-}-\gamma_{ss}\right)&=&f_Q \left(
\langle\langle\cos\Delta\phi_{\rm balance}\rangle\rangle+\frac{v_{2,c}}{v_2}-\frac{v_{2,s}}{v_2}\right),\\
\nonumber
\frac{3M}{4v_2}\gamma_{ss}&=&f_P\frac{\left(1+\langle\langle\cos\Delta\phi_{\rm balance}\rangle\rangle\right)}{2}.
\end{eqnarray}
The quantities on the left-hand-sides of the expressions above are plotted in Fig. \ref{fig:results}. The multiplicity $M$ was taken by assuming that the detector covered 2 units of pseudo-rapidity with the multiplicity being equal to 1.85 times the number of participants. The factor 1.85 comes from a PHOBOS analysis that showed the multiplicity scaling with the number of participants to the 10\% level \cite{Back:2006yw}.  The number of participants depends on the centrality and its extraction is modestly model dependent.  Values of $v_2$ were also taken from the event-plane analysis of STAR \cite{Abelev:2008ed}.

To estimate the first term on the right-hand sides of Eq. (\ref{eq:scaled}) one must estimate the fraction of balancing charge falling within the rapidity range of the detector $f_Q$, and the width of the charge balance function $\langle\langle\cos\Delta\phi_{\rm balance}\rangle\rangle$. The fraction $f_Q$ of balancing charge found within the 2 units of rapidity covered by STAR can in principle be determined from balance function analyses. These analyses showed that the integrated strength of the balance function describes the chance of finding a balancing charge. This probability tended to be in the neighborhood of 0.5 for non-identified particles. However, some of this reduction was due to the non-ideal efficiency of the detector or the $p_t$ cutoffs. Since the quantity $M$ used to scale the correlations was corrected for efficiency and for $p_t$ cutoffs, the fraction $f_Q$ should only refer to the fraction of balance charge that lay outside the rapidity acceptance of STAR, roughly $\pm 1$ units. Thus, one would expect $f_Q$ to be approximately 0.75. Balance functions from STAR in term of the relative azimuthal angle have appeared in preliminary form and it appears that the widths have $\langle\langle\cos\Delta\phi_{\rm balance}\rangle\rangle$ are near 0.5 within a few tens of percent. Combining with an estimate of $f_Q$, it would not be surprising if the first term, $f_Q\langle\langle\cos\Delta\phi_{\rm balance}\rangle\rangle$, in Eq. (\ref{eq:scaled}) is approximately 1/3, which would explain a bit more than half of the observed correlation in Fig. \ref{fig:results}. Since $v_{2,s}$ and $v_{2,c}$ have not been evaluated, even in preliminary form, one can only speculate as to whether those contributions could explain the remainder of the measured correlation. One does expect $v_{2,c}$ to be positive and $v_{2,s}$ to be negative, hence, they should help push the right-hand side toward the data, but with an unknown amount. Thus, one can only conclude that the majority of the correlation $\gamma_{+-}-\gamma_{ss}$ seems to derive from combining charge conservation with elliptic flow, but the possibility remains that this source could fall short of the total observed correlation by a few tens of percent.

If the measurement of $\gamma_{+-}-\gamma_{ss}$ turns out to be too large to be described by folding local charge conservation with elliptic flow, it would demand considering other sources of correlation. As shown in Sec. \ref{sec:other}, correlations deriving from the average electric field in the initial participant region being non-zero, tend to be too small to account for more than a percent of the signal. Even though the fields might give opposite impulses to positive and negative particles of the order of 5 MeV/$c$, those impulses are only given to those charges which exist in the first few fm/$c$, and then tend to be muted by the particles being reabsorbed and re-thermalized in the medium. The same considerations apply to electric fields generated by coupling to parity fluctuations in the QCD sector. If the generated electric fields are 10\% of the magnetic field strength, the imparted momenta are also on the order of 5 MeV/$c$. Again, the considerations of having only a fraction of charge affected, and of having particles re-thermalized, would make such a signal far too small. Thus, if coupling electro-magnetic and QCD fields through the anomalous coupling is to account for a significant fraction of the observed correlation, the generated electric field would have to be nearly as strong as the driving magnetic field. 

The correlations $\gamma_{+-}$ and $\gamma_{ss}$ represent moments of the full differential angular correlations which would be functions of $\phi_\alpha$ and $\phi_\beta$. In fact, the relation for $\gamma_{+-}-\gamma_{ss}$ in Eq. (\ref{eq:scaled}) is more a consistency check than a test of charge conservation. The most telling way to distinguish correlations is to view the correlations in the most differential form possible, given the limited statistics. For instance, the balance function can be evaluated in terms of the two angles,
\begin{equation}
B(\phi_\alpha,\phi_\beta)\equiv\frac{N_{+-}(\phi_\alpha,\phi_\beta)+N_{-+}(\phi_\alpha,\phi_\beta)-N_{++}(\phi_\alpha,\phi_\beta)-N_{--}(\phi_\alpha,\phi_\beta)}
{N_+(\phi_\alpha)+N_-(\phi_\alpha)}.
\end{equation}
This observable can be thought of as the conditional distribution for seeing an extra particle of opposite sign at $\phi_2$, given the observation of a first particle at $\phi_1$, and integrating over $\phi_2$ should yield unity for perfect acceptance and efficiency (also assuming equal numbers of positives and negatives). Subtractions of like-sign and opposite-sign correlations would carry the same information. Local charge conservation leads to a peak for $\phi_1$ near $\phi_2$. For the parity violation hypothesis, or for the electric field acting on a fluctuating density distribution, one would expect the balance function to have a peak for $\phi_1-\phi_2\sim$ 180 degrees, and the peak should be more pronounced for $\phi_1$ near $\pm 90$ degrees. Preliminary charge balance functions have not shown such a peak for $\Delta\phi\sim 180^\circ$, but analyses have not yet been performed where $\phi_1$ is constrained to be near $\pm 90^\circ$. Furthermore, one can help discriminate sources of correlation by binning according to relative rapidity. If the source of the correlation is charge conservation plus elliptic flow, the peak should be confined in relative rapidity to one unit, the width of the measured balance function in relative rapidity. Since correlations from fluctuating initial conditions extend over multiple units of rapidity, one would expect the correlation due to fluctuating initial conditions combined with electric fields from the incoming ions, described in Sec. \ref{sec:other}, to extend over a wide range of rapidity. Assuming that the direction of the color magnetic flux also extends over the entirety of a flux tube, one would also expect parity fluctuation induced correlations to also extend well beyond one unit of rapidity. The high statistics of recent data sets at RHIC indeed make it possible to perform such differential analyses.

The right-hand side of Eq. (\ref{eq:scaled}) for the scaled same-sign correlation can also only be estimated at the current time. The fraction $f_P$ describes the fraction of the momentum responsible for $v_2$ that falls within the same rapidity range must be less than unity, and if the rapidity range is dominated by final thermal motion, one might expect, within a few tens of percent,  $f_P\sim 3/4$.  If $\langle\langle\cos\Delta\phi_{\rm balance}\rangle\rangle\sim 0.5$, the right-hand side would be approximately 1/2, which would account for only half the signal shown in Fig. \ref{fig:results}. The estimate for $\gamma_{ss}$ from momentum conservation in Eq. (\ref{eq:scaled}) was based on an a model where all the particles had the same $p_t$. As emphasized in Sec. \ref{sec:momentumconservation}, evaluating the $p_t$ weighted correlation as defined in Eq. (\ref{eq:gammaprime}) should reduce the model dependence and give a better feel for how much of the observed correlation comes from momentum conservation combined with elliptic flow. It is possible that other sources of correlation are responsible for a significant fraction of the signal. For example, the quenching of jets, or mini-jets, might have a reaction-plane dependence. 

If the same-sign correlation is indeed largely driven by momentum conservation combined with anisotropic flow, this correlation might provide important insight into dynamics and perhaps even bulk properties of the matter. Gavin has proposed using $p_t$ correlations binned by relative rapidity as a means to gain insight into viscosity or heat conductivity \cite{Gavin:2006xd}. However, it is difficult to understand the degree to which jets interfere with such interpretations, as even soft-sector particles may have originated from quenched jets or mini-jets. In contrast, the quantity $\gamma_{ss}^\prime$ is only sensitive to the part of the momentum correlations that contribute to the elliptic anisotropy. Thus, if $\gamma_{ss}'$ were analyzed as a function of the relative rapidity, one might gain a more robust insight into transport coefficients.

Returning to the motivating suggestion that parity fluctuations drive these observables, it now seems that the bulk of the observed correlation $\gamma_{+-}-\gamma_{ss}$ can be explained by the combined effects of local charge conservation and elliptic flow. The first term in Eq. (\ref{eq:gamma_chargeconservation_general}) would seem to explain half or more of the contribution, and after accounting for the additional terms, this has the potential to explain nearly all the signal. Further analysis of charge balance functions should show the degree to which the three terms in Eq. (\ref{eq:gamma_chargeconservation_general}) contribute individually to the effect. The same-sign correlation, which by itself does not promote the idea of parity fluctuations, might be similarly explained by momentum conservation coupled to elliptic flow. However, until $p_t$ weighted correlations are constructed, more quantitative conclusions are unwarranted.

Fluctuating initial conditions combined with the electric fields of the spectator nuclei in the previous section also provide correlations of the same sign as those observed. These effects have much in common with the effect one would expect from parity fluctuations. The correlations would extend multiple units of rapidity and would affect opposite-sign and same-sign correlations in the opposite direction. However, the estimates of the previous section suggest that such contributions might are too small to be observable, even though the effective electric fields were of the order of 10\% of the strength of the magnetic fields. Given that the electric field generated from the magnetic field coupling to parity-odd term in the QCD fields might easily be less than 10\% of the magnetic field, this underscores that explaining the observed correlations through parity fluctuations would require surprisingly large couplings of the parity odd ${\bf E}\cdot{\bf B}$ terms in the QCD and electromagnetic sectors.

\section*{Acknowledgments}
Support was provided by the U.S. Department of Energy, Grant No. DE-FG02-03ER41259. The author thanks Gary Westfall, Hui Wang, Terence Tarnowsky, Fuqiang Wang, Gang Wang and Sergei Voloshin for insightful discussions of STAR data.


\begin{thebibliography}{99}
\bibitem{STARparity} B.I. Abelev et al. (STAR Collaboration), 0909.1717;
B.I. Abelev et al., Phys. Rev. Lett. {\bf 103}, 251601 (2009).
\bibitem{Kharzeev:2004ey}
  D.~Kharzeev,
  Phys.\ Lett.\  B {\bf 633}, 260 (2006).
\bibitem{Kharzeev:2009mf}
  D.~E.~Kharzeev,
  Nucl.\ Phys.\  A {\bf 830}, 543C (2009),
  [arXiv:0908.0314 [hep-ph]].
\bibitem{Adams:2003kg}
  J.~Adams {\it et al.}  [STAR Collaboration],
  Phys.\ Rev.\ Lett.\  {\bf 90}, 172301 (2003).
  [arXiv:nucl-ex/0301014].

\bibitem{Bass:2000az}
  S.~A.~Bass, P.~Danielewicz and S.~Pratt,
  Phys.\ Rev.\ Lett.\  {\bf 85}, 2689 (2000).
  
\bibitem{Cheng:2004zy}
  S.~Cheng {\it et al.},
  Phys.\ Rev.\  C {\bf 69}, 054906 (2004).

\bibitem{Abelev:2008ed}
  B.~I.~Abelev {\it et al.}  [STAR Collaboration],
  Phys.\ Rev.\  C {\bf 77}, 054901 (2008).

\bibitem{Westfall:PC}
G.~D. Westfall, private communication.


\bibitem{Back:2006yw}
  B.~B.~Back {\it et al.}  [PHOBOS Collaboration],
  Phys.\ Rev.\  C {\bf 74}, 021902 (2006).

\bibitem{Gavin:2006xd}
  S.~Gavin and M.~Abdel-Aziz,
  Phys.\ Rev.\ Lett.\  {\bf 97}, 162302 (2006).
\end{thebibliography}
\end{document}